**Suppression of Metallic Transport in Nitrogen-rich Two-Dimensional Transition Metal Nitrides**

*Hongze Gao*[1], *Da Zhou*[2], *Nguyen Tuan Hung*[3], *Chengdong Wang*[4], *Zifan Wang*[1], *Ruiqi Lu*[5], *Yuxuan Cosmi Lin*[4,6], *Jun Cao*[1], *Michael Geiwitz*[7], *Gabriel Natale*[7], *Kenneth S. Burch*[7], *Xiaofeng Qian*[4], *Riichiro Saito*[8,9], *Mauricio Terrones*[2,10,11], *and Xi Ling*[1,5,*]

[1]Department of Chemistry, Boston University

590 Commonwealth Ave., Boston, MA, 02215, U.S.

[2]Department of Physics, the Pennsylvania State University

104 Davey Laboratory, University Park, PA, 16802, U.S.

[3]Department of Materials Science and Engineering, National Taiwan University

Taipei 10617, Taiwan

[4]Department of Materials Science and Engineering, Texas A&M University

575 Ross St., College Station, TX, 77843, U.S.

[5]Division of Materials Science and Engineering, Boston University

15 St Mary's St., Boston, MA, 02215, U.S.

[6]Department of Electrical and Computer Engineering, Texas A&M University

188 Bizzell St, College Station, TX 77843, U.S.

[7]Department of Physics, Boston College

140 Commonwealth Ave., Chestnut Hill, MA, 02467, U.S.

[8]Department of Physics, Tohoku University

Sendai 980-8578, Japan

[9]Department of Physics, National Taiwan Normal University

Taipei 11677, Taiwan

[10]Department of Chemistry, the Pennsylvania State University

University Park, PA, 16802, U.S.

[11]Department of Materials Science and Engineering, the Pennsylvania State University

University Park, PA, 16802, U.S.

*Corresponding author: xiling@bu.edu

## Abstract

The recent experimental realization of two-dimensional (2D) transition metal nitrides (TMNs, *e.g.*, $Mo_5N_6$, δ-MoN, and $W_5N_6$) opens new opportunities for exploring their fundamental physical properties at the two-dimensional limit. In this work, we propose a unified picture of transport phenomena in the nitrogen-rich 2D $W_5N_6$ and $Mo_5N_6$, and the stoichiometric 2D δ-MoN based on several observations and first-principles calculations. Temperature coefficient of resistance (TCR) and magnetoresistance (MR) from Hall measurements consistently suggest disorder-induced transport mechanism at low temperatures (10-30 K). Notably, we observe a transition from metal to semimetal driven by the variation of nitrogen content in TMNs, supported by the suppressed density of states at the Fermi energy in nitrogen-rich TMNs (e.g. $Mo_5N_6$) from first-principle calculations. Carrier density calculations of bulk TMNs and 2D TMNs with -NH termination groups further reveal the switching of majority carrier type of $Mo_5N_6$ at reduced thickness, which is in great agreement with Hall measurement results. Our findings demonstrate that high nitrogen content in metallic molybdenum nitrides can induce the transition to a semimetallic phase at the 2D limit, shedding light on both the fundamental aspects of these materials and directions in future material design.

# I. INTRODUCTION

The past two decades have witnessed significant interest in two-dimensional (2D) crystals comprising single or few atomic layers, driven by their high specific surface area, pronounced quantum confinement effect, and broken symmetry at reduced dimensions [1–4]. These fundamental alterations have sparked rapid advancements in the field, unveiling novel physical properties of 2D materials distinct from their bulk counterparts [5,6], and showcasing their vast potential in applications across nanoelectronics [7,8], quantum information [9,10], and energy technologies [11–13]. Although research has been largely centered on van der Waals (vdW) 2D materials such as graphene and transition metal chalcogenides (TMDs) because of their synthesis accessibility, recent advances have enabled the realization of non-vdW 2D materials such as 2D transition metal carbides, nitrides, or carbonitrides (MXenes) [14,15], 2D metals [16], and 2D metal oxides [17], creating new opportunities to investigate their properties at the 2D limit. For example, Cao *et al.* reported an atomic substitution approach to chemically convert vdW layered TMDs into ultrathin non-vdW transition metal nitrides (TMNs) crystals via a nitridation reaction [18], where the thickness of the resulting TMNs such as $Mo_5N_6$ and $W_5N_6$ can be tuned and precisely controlled by the number of layers in the TMDs precursors. Various phases of TMNs were also realized by varying the reaction temperature, as evidenced by their distinct Raman signatures [19–21]. In addition, Chin *et al.* used a vapor–liquid–solid growth process to synthesize single-unit-cell thick $W_5N_6$ crystals on sapphire substrates with high chemical resistance [22]. Bulk TMNs are a class of highly conductive ultrahigh-temperature ceramics with densely packed hexagonal crystal structures, featuring strong covalent nitrogen bonds and metal-nitrogen interactions, which results in exceptional hardness and chemical and thermal stability [23]. Moreover, multiple stable phases with distinct stoichiometric ratio between metal and nitrogen render this class of materials rich tunability [24–26]. Taking molybdenum nitrides as an example, Fig. 1a-d illustrates the crystal structures of $Mo_5N_6$ and δ-MoN. The $Mo_5N_6$ structure can be viewed as a δ-MoN framework with periodic molybdenum cation vacancies in one-sixth of the Mo sites. In this work, we refer to the $Mo_5N_6$ and $W_5N_6$ phases as nitrogen-rich structures, in contrast to the stoichiometric δ-MoN phase.

Access to this new class of non-vdW 2D crystals enables systematic exploration of their physical properties and their potential for advancing electronic applications. It has been reported that $Mo_5N_6$ and δ-MoN both exhibit high electrical conductivity, with δ-MoN (3126 S/cm) approximately ten folds higher than $Mo_5N_6$ (229.6 S/cm) [19]. Chin *et al.* further demonstrated the semimetallic electrical behavior from field effect transistor (FET) transconductance measurements and robustness of $W_5N_6$, highlighting its promise as a 2D contact material [22]. In addition, Gao *et al.* showed that the resistivity of $W_5N_6$ of around 161.1 S/cm is preserved upon downscaling to a thickness of 2.9 nm [20], highlighting the potential of TMNs produced via the atomic substitution approach as future interconnect materials for advanced node technologies. Nevertheless, current studies on the electrical properties of 2D TMNs are limited to the basic characterizations of the electrical conductivity, and a mechanistic understanding of charge transport in 2D TMNs remains lacking. For example, although $W_5N_6$, $Mo_5N_6$, and δ-MoN have been reported to exhibit linear *I–V* characteristics at room temperature [18–20], such observations alone are insufficient to unambiguously establish their metallic nature. Complementary evidence, such as temperature-dependent transport measurements, is still required. Moreover, the physical origin of the pronounced differences in electrical conductivity among these TMNs remains unclear, which is counterintuitive in view of the similar electronic structures of Mo and W atoms. Hall measurements, which directly provide carrier density and mobility information, are therefore essential for elucidating the transport mechanisms in these materials. In addition, owing to the non-vdW crystal structure and the presence of unsaturated, dangling-bond-rich surfaces, extrinsic surface termination species are expected to form and can strongly affect the electronic properties at 2D limit [27]. However, their role has not yet been investigated for TMNs.

In this work, we elucidate the charge transport characteristics of 2D TMNs including $Mo_5N_6$, δ-MoN and $W_5N_6$. By combining temperature coefficient of resistance (TCR) and magnetoresistance (MR) from Hall measurements, we identify the dominant charge transport mechanisms across different temperature regimes. At high temperatures (230-300 K), notably, we observe semimetallic transport behavior in the $Mo_5N_6$ and $W_5N_6$ phases evidenced by TCR measurement, which is attributed to the modification of band structure compared to δ-MoN due to the rich nitrogen content. Band structure calculations show that nitrogen-rich $Mo_5N_6$ exhibit suppressed density of states (DOS) around the Fermi energy ($E_F$) compared to the stoichiometric δ-MoN, in good agreement with our conductivity measurements. These findings demonstrate that high nitrogen content in metallic TMNs can drive a transition from metal to semimetal at the 2D limit, providing insight into both the fundamental transport physics of TMNs and guiding principles for future materials design.

## II. EXPERIMENTAL DETAILS

### A. Materials synthesis and characterizations

2D $Mo_5N_6$ and $W_5N_6$ are obtained through the atomic substitution approach where nitridation reactions are performed on 2D $MoS_2$ and $WSe_2$ flakes exfoliated from bulk crystals, respectively [18,20]. The reaction is conducted in a quartz tube at ~750 °C with $NH_3$ gas as nitrogen source [18,19,28–30]. 2D δ-MoN is synthesized by annealing the $Mo_5N_6$ flakes in pure Ar atmosphere at 800 °C for 30 minutes for complete phase transition. Typical optical image of the precursor $MoS_2$ flake and converted $Mo_5N_6$ flake are shown in Fig. 1e and 1f. More optical images of δ-MoN and $W_5N_6$ are shown in Fig. S1. The smooth surface of the TMNs is evidenced by the topography measurements using atomic force microscopy (AFM), as shown in Fig. 1g. The thickness of the flakes is extracted from AFM topography measurements acquired using a Bruker Dimension system. Raman spectra as presented in Fig. 1h [19,20], where the TMNs show distinct Raman spectra that helps identify the synthesized 2D flakes.

### B. Device fabrication and electric transport measurements

All electrodes were fabricated inside a cleanroom in a glovebox [31]. The electrode patterns were done using a bilayer photoresist (LOR1A/S1805) and laser mask writer (Heidelberg Instruments). Then, Cr/Au (5/45 nm) layers were deposited using e-beam evaporation as contact electrodes to the 2D flakes. Lift-off of extra metal is done using remover PG (MicroChem).

Electrical transport properties were measured using Quantum Design Dynacool Physical Properties Measurement System (PPMS) with a DC excitation current. Resistance measurements were done using a four-point probe method. A standard six-probe Hall bar geometry was used for the Hall measurements. Magnetic field is applied perpendicular to the sample plane (H//c). To eliminate the error caused by misalignment of electrodes, $R_{xx}$ and $R_{xy}$ are symmetrized and asymmetrized through the equations below, respectively:

$$R_{xx_{symmetrized}}(B) = \frac{R_{xx_{raw}}(B) + R_{xx_{raw}}(-B)}{2} \tag{1}$$

$$R_{xy_{asymmetrized}}(B) = \frac{R_{xy_{raw}}(B) - R_{xy_{raw}}(-B)}{2} \tag{2}$$

MR is further defined as:

$$\frac{R_{xx}(B) - R_{xx}(0)}{R_{xx}(0)} \times 100\% \tag{3}$$

$R_{Hall}$ is defined as the asymmetrized $R_{xy}$. Carrier concentration and Hall mobility is extracted from $R_{Hall}$ using a single band model:

$$n_{2D} = \frac{1}{q} \cdot \frac{dB}{dR_{Hall}} \tag{4}$$

$$\mu = \frac{1}{R_{sheet}} \cdot \frac{dR_{Hall}}{dB} \cdot \frac{q}{e} \tag{5}$$

Here, $q = -e$ for electrons and $+e$ for holes ($e$ is the unit charge), $R_{sheet}$ is the sheet resistance extracted from I-V measurements. We further convert the 2D carrier density into 3D carrier density considering the thickness of the samples using the following equation:

$$n_{3D} = \frac{n_{2D}}{d} \tag{6}$$

where d is the sample thickness extracted from AFM.

### C. Band structure calculations

The energy band structures and the energy density of states (DOS) of four structures, including close-packed molybdenum nitride (δ-MoN), close-packed molybdenum nitride with periodic Mo vacancies ($Mo_5N_6$), close-packed tungsten nitride (WN), and close-packed tungsten nitride with periodic W vacancies ($W_5N_6$), are calculated by using the density functional theory (DFT) with Quantum ESPRESSO [32]. The hexagonal unit-cells of δ-MoN and WN consists of 12 atoms, while that of $Mo_5N_6$ and $W_5N_6$ consists of 11 atoms, as shown in Fig. S2. All atomic position and lattice vectors are fully optimized by using the BFGS method with convergence criteria for force and stress set to 1.0×10-4 Ry/a.u. and 0.005 GPa, respectively [33]. The optimized lattice vectors (*a*; *c*) are (4.927 Å; 5.643 Å), (4.888 Å; 5.698 Å), (4.852 Å; 5.538 Å), and (4.828 Å, 5.549 Å) for δ-MoN, WN, $Mo_5N_6$ and $W_5N_6$, respectively. A cut-off energy of 60 Ry is selected for the plane-wave basis. Here, we adopt the full-relativistic pseudopotential with projector-augmented-wave method (PAW) [34], including Mo.rel-pz-spn-kjpaw_psl.0.2.UPF, W.rel-pz-spn-kjpaw_psl.1.0.0.UPF, and N.rel-pz-n-kjpaw_psl.0.1.UPF for the Mo, W and N atoms, respectively. The spin-orbit coupling (SOC) is taken into account in the calculations. The k-grid of 8×8×8 is used for the band structure calculations, while a dense k-grid of 20×20×20 with the tetrahedron method is used for the DOS calculations [33].

### D. Carrier density calculations

Carrier density was calculated using first-principles DFT as implemented in the Vienna Ab initio Simulation Package (VASP). The projector-augmented wave (PAW) method was employed to describe the interaction between ion cores and valence electrons, together with a plane-wave basis set with an energy cutoff of 520 eV. Structural relaxations including both atomic positions and lattice parameters were carried out using the conjugate-gradient algorithm until the total energy and atomic forces converged to within $10^{-6}$ eV and 0.01 eV/Å, respectively. For the Brillouin zone sampling, a Monkhorst-Pack *k*-point grid of 12×12×12 was used for the Brillouin zone sampling in bulk structural relaxation, while a grid of 12×12×1 for 2D $Mo_5N_6$. Self-consistent calculations for the carrier density were then performed using denser *k*-point grids of 30×30×10 and 30×30×1 for bulk and 2D system, respectively, with SOC taken into account.

The carrier density was calculated by integrating the *k*-point dependent electronic occupations over the Brillouin zone using a Fermi–Dirac distribution with a smearing factor of 0.2 eV. The electron contribution was obtained from the occupied fraction of the conduction bands above the Fermi level, while the hole contribution was obtained from the unoccupied fraction of valence bands, respectively, as shown below.

$$n_e = \int_{E_F}^{\infty} f(E)\, g(E)\, dE \tag{7}$$

$$n_h = \int_{-\infty}^{E_F} [1 - f(E)]\, g(E)\, dE \tag{8}$$

## III. Results and discussions

### A. Temperature coefficient of resistance

Fig. 2a presents the resistance measurements of the three TMNs from 4 K to 300 K, with resistance values normalized to the corresponding measurement at 300 K. We also plot the TCR (defined as $\frac{1}{R}\cdot\frac{dR}{dT}$) in Fig. S3 for more clear visualization. Notably, the TCR varies in two aspects: (1) TCR of each individual TMNs varies with temperature, and (2) TMNs exhibit distinct TCR from each other in the same temperature regime. To understand the data, we divide the entire temperature range into three phases based on different fitting models: 10-25 K (phase I, labelled with light blue colour), 25-230 K (phase II, labelled with light yellow colour), and 230-300 K (phase III, labelled with light red colour). In phase I, all TMNs show negative TCR, which is a typical behaviour of insulating materials. In phase III, nitrogen-rich $W_5N_6$ and $Mo_5N_6$ present a negative TCR with smaller magnitude compared to phase I, while δ-MoN exhibits a positive TCR, which is a typical behaviour of metallic materials. In Phase II, $Mo_5N_6$ and $W_5N_6$ exhibit changes only in the magnitude of the TCR, while the sign of the TCR remains unchanged. In contrast, δ-MoN shows changes in both the sign and magnitude of the TCR. In this work, we use the sign of TCR [35,36], despite the values of electrical conductivity. Hereby, our results show two sets of transitions: (1) In δ-MoN, the material transits from an insulating state in phase I to a metallic state in phase III with temperature rises across ~30 K; (2) At higher temperature (phase III), a comparison between the Nitrogen-rich $Mo_5N_6$ and stoichiometric δ-MoN shows metal-semimetal transition, which is later attributed to the variation in nitrogen content and further supported by first-principle calculations (details discussed in section C). The distinct values and/or signs of TCR in phase I and III indicate that different mechanisms are dominating the charge transport characteristics in 2D TMNs within each phase, and hence different models are required for data interpretation. As for phase II, we consider this as an intermediate phase, where the mechanism in phase I and phase III contribute simultaneously to the physical picture, imposing technical difficulties to quantitatively analyze the data. Therefore, the majority of discussion will focus on phase I and phase III individually.

To understand the negative TCR in phase I, we reorganized the data in an R-T$^{-1/3}$ plot with resistance in logarithm scale in Fig. 2c. The colour labels inherit from Fig. 2a and resistance values are normalized to the corresponding value at 100 K. The data of all 2D TMNs can be fitted into the variable range hopping (VRH) model (dashed line in Fig. 2c) with minimal residues ($R^2 \geq 0.995$) [37–39]:

$$R(T) = R_0 \exp\left[\left(\frac{T_0}{T}\right)^{\frac{1}{3}}\right] \tag{9}$$

Where $T$ stands for temperature, $R(T)$ is the resistance at temperature $T$, $R_0$ and $T_0$ are constants. The $\frac{1}{3}$ power relationship suggests that the hopping takes place within a 2D system [37], which matches with the 2D nature of the TMN samples. The VRH model is applied to describe the current flow in Anderson insulating systems [37–39], where direct charge transport is interrupted due to the disorder crystal structure. Considering that the three TMNs are prepared using a similar atomic substitution approach, we attribute the VRH mechanism in phase I of all TMNs to the disorders introduced during the synthesis process.

To interpret the distinct TCR of TMNs in phase III, we plotted the normalized conductance ($G/G_{300\,K}$) as a function of $\frac{1}{T}$ as shown in Fig. 2b. Different models are used to fit the experimental data. For $W_5N_6$ and $Mo_5N_6$, an Arrhenius-like correlation fits the data well with minimal residual:

$$G(T) = G_0 \cdot \exp\left(-\frac{E_a}{kT}\right) \tag{10}$$

Here $G(T)$ is the conductance at temperature $T$, $G_0$ *is* a constant, $k$ is the Boltzmann constant, and $E_a$ is an activation energy which can be extracted from the fitting result (plotted in Fig. S4). The extracted $E_a$ values are fairly small (< 10 meV) compared to the bandgap energy of common semiconductors [5,6], which are usually in the range from 0.5 to 2 eV. Moreover, we found that the $E_a$ value decreases with thicknesses (Fig. S4), which can be attributed to enhance quantum confinement effect with decreased thickness [28]. Hereby in phase III, the presence of $E_a$ identifies $W_5N_6$ and $Mo_5N_6$ as semimetals [35]. In contrast, the results of δ-MoN can be well fitted using the following equation:

$$G(T) = G_0 \cdot \frac{1}{T} \tag{11}$$

In this case, the temperature dependence of the conductance is primarily governed by increased scattering between charge carriers and phonons, leading to reduced carrier mobility at elevated temperatures [40]. This is a typical metallic property. The fittings to all TMNs in phase III are presented by dashed lines in Fig. 2b with minimal residue ($R^2 \geq 0.996$). The minimal discrepancy between fitting results and experimental data validates the semimetallic nature of $W_5N_6$ and $Mo_5N_6$, and the metallic nature of δ-MoN in phase III.

### B. Hall measurements

To obtain more insights on the charge transport behaviour in 2D TMNs, we further perform Hall measurements on the samples. Fig. 3a shows a typical optical image of the measured 2D TMNs flake with a Hall bar electrodes geometry. Magnetic field is applied perpendicular to the sample plane as labelled in the image. Fig. 3b shows typical I-V characteristic curves of TMNs measured with four-point probe configurations at room temperature, where linear trend is observed. δ-MoN exhibits higher current than $Mo_5N_6$ and $W_5N_6$ with similar channel dimensions, suggesting higher metallicity of δ-MoN. We performed Hall measurements on the samples at low temperatures down to 2 K and extracted MR and $R_{Hall}$ data, with details discussed in EXPERIMENTAL DETAILS section. Fig. 3c presents a typical $R_{Hall}$-$B$ plot of 2D TMNs measured on a 6.5-nm-thick $Mo_5N_6$ flake from 2 K to 100 K, where a linear trend with negligible change in the slope is observed. The constant slope suggests little variation of carrier concentration throughout this temperature range. When the temperature increases beyond 100 K, a significant thermal noise starts to interfere with the measurements, resulting in a low signal-noise-ratio (Fig. S5). We summarize the key parameters of the TMNs extracted from Hall measurements in Table 1.

As shown in Table 1, a positive correlation is observed between the conductivity and carrier density of $Mo_5N_6$, where the 6.5-nm-thick sample possess highest carrier density and electrical conductivity. All TMNs samples exhibit fairly high carrier density ($10^{22}$~$10^{23}$ $cm^{-3}$). The high carrier density aligns with their high conductivity and linear I-V characteristics, which have been reported on both 2D and bulk TMNs [19,41]. Such high carrier density is comparable to that of many thin film TMNs such as NbN and TiN [42–44]. We also calculated the ratio $n_{eff}/n$ from the experimental results, where $n_{eff}$ is the effective charge carrier density extracted from Hall measurement, and n is the total number of delocalized *4d* electrons. Taking $Mo_5N_6$ as an example, the formula cell of $Mo_5N_6$ contains 5 Mo and 6 N atoms, where each Mo provides 6 electrons to the N *2p* orbitals and Mo *4d* conduction bands. 12 delocalized electrons in the *4d* conduction bands per unit cell (i.e. "5×6 $e^-$ provided by Mo" – "6×3 $e^-$ to N *2p* orbital" = 12 delocalized electrons) results in a free electron density n = $9.81\times10^{22}$ $cm^{-3}$ and $n_{eff}/n$ = 0.18. The small ratio is attributed to the semimetal characteristics of $Mo_5N_6$. In contrast, the $n_{eff}/n$ value is 4.00 for δ-MoN, which aligns with its metallic TCR behavior.

Moreover, we also noticed that the charge carrier type varies with the thickness of $Mo_5N_6$. As the thickness of $Mo_5N_6$ increases, the major charge carrier switched from electrons to holes. We attribute this unusual behavior to surface termination groups that often play important roles in materials properties at nanometer scale [27]. We performed X-ray photoelectron spectroscopy (XPS) on a 2-nm thick $Mo_5N_6$ flake to characterize the surface termination group. As shown in Fig, S6a, two peaks associated to N 1s orbital are observed with binding energy of 398.5 eV and 397.2 eV, respectively. The 397.2 eV peak is attributed to Mo-N bonding as we observed in our previous work [18], whereas the 398.5 eV peak is attributed to N-H bonding [45], which is potentially introduced by $NH_3$ gas used for the nitridation reaction. The addition of H atom as surface termination donates one extra electron to the crystal, causing n-type doping effect. Such doping effect is predominant at reduced dimension, and eventually switch the majority charge carrier type from holes in bulk $Mo_5N_6$ to electrons in 2D $Mo_5N_6$. More details are discussed in the calculation results in Section C.

Apart from the $R_{Hall}$ data, we also measured the MR of TMNs, with typical results of each TMNs shown in Fig. 3d-3f. A small negative MR of < 0.25% at 3.0 T is observed below 10 K for all three TMNs, which deviates from the normal situation where MR is typically positive MR caused by Lorentzian force [46,47]. As temperature exceeds 10 K, the magnitude of MR rapidly decreases to the noise level for $Mo_5N_6$ and δ-MoN, while it changes to a small positive MR for $W_5N_6$. Multiple mechanisms have been reported with the potential to trigger a negative MR in Hall measurements, such as semimetallic band structure [48], high pressure [49,50], strain [51], and defect-induced weak localization effect [52]. To identify the origin of the negative MR, we compare the experimental conditions required for different proposed mechanisms. Although a negative TCR can in some cases be observed in semimetallic systems or in strained materials, the clear VRH behavior observed in the same temperature range demonstrates that charge transport is dominated by localized states rather than band-like carriers. Therefore, the observed negative MR is unlikely to originate from semimetallic band transport or strain-induced band-structure modification. In addition, pressure-induced mechanisms can be excluded, since all magnetotransport measurements were performed under high vacuum. Considering the presence of disorder-induced VRH mechanism evidenced by TCR in phase I (Fig. 2c), we attribute the negative MR to the weak localization effect [52], which share the same origin (*i.e.*, disorders) as the VRH mechanism. Similar phenomena have also been reported on Cu thin films [53], carbon fibers [54], and other 2D materials [37,55].

### C. Band structure calculations

To investigate the origin of lower metallicity of $Mo_5N_6$ and $W_5N_6$, we perform electronic band structure and DOS calculation on four structures, including δ-MoN, $Mo_5N_6$, WN, and $W_5N_6$. From the calculated energy band structure shown in Fig. 4, there is no obvious gap around $E_F$ in all four structures, suggesting a metallic characteristic of the crystals. Notably, upon the introduction of cation vacancies, DOS around $E_F$ is significantly suppressed for both molybdenum and tungsten nitrides. As shown in Fig. 4c, DOS at $E_F$ decreased from over 6 state/eV/u.c. for δ-MoN to less than 1 state/eV/u.c. for $Mo_5N_6$. Similar trend is observed from WN to $W_5N_6$. Therefore, a lower carrier concentration is expected in the nitrogen-tich $Mo_5N_6$ and $W_5N_6$ compared to the stoichiometric δ-MoN and WN. The presence of a small yet finite DOS at $E_F$ points to semimetallic behaviour in the nitrogen-rich $Mo_5N_6$ and $W_5N_6$. Moreover, the existence of a van Hove singularity in the valence band near $E_F$ (~0.1 eV) implies that the system is close to a Lifshitz transition, highlighting the sensitivity of its electronic structure to external tuning [56]. Although the calculation of $W_5N_6$ and WN cannot yet be experimentally verified due to the absence of synthesis method for WN, the calculated differences between $Mo_5N_6$ and δ-MoN are in good agreement with their experimentally observed TCR behavior. We further calculated the carrier density of TMNs and summarized the results in Table S1, where high carrier densities of ~$10^{21}$ $cm^{-3}$ have been predicted. The high carrier density matches with the metallic property (i.e. high conductivity and high Hall carrier concentration) of TMNs. Moreover, we also noticed that the H-terminated 2D $Mo_5N_6$ exhibits higher electron density compared to the bulk $Mo_5N_6$, which is attributed to the n-doping of H termination groups. This is consistent with the experimental observation from Hall effect.

## IV. Conclusions

In this work, we investigated the charge-transport properties of 2D TMNs including nitrogen-rich $W_5N_6$ and $Mo_5N_6$, and stoichiometric δ-MoN. At low temperatures, all samples exhibit disorder-induced VRH transport, as evidenced by the TCR measurements, together with a weak-localization contribution manifested by negative MR. At higher temperatures, a clear metallic and semimetallic transport transition is observed when comparing the stoichiometric δ-MoN with the nitrogen-rich $Mo_5N_6$, demonstrating that the introduction of cation vacancies suppresses metallic transport in 2D TMNs. In addition, δ-MoN exhibits a temperature-driven crossover from insulating to metallic transport behavior, which is attributed to a change in the dominant transport mechanism from VRH at low temperatures to phonon-limited band transport at elevated temperatures. Moreover, all TMNs exhibit high carrier densities, and a surface-termination-induced carrier-type switching from p-type in bulk $Mo_5N_6$ to n-type in ultrathin $Mo_5N_6$ is observed. Consistent with the experimental results, band-structure calculations reveal a pronounced suppression of the density of states at the Fermi energy in nitrogen-rich TMNs compared with stoichiometric δ-MoN, accounting for the reduced metallicity of $Mo_5N_6$. Carrier-density calculations further reproduce the thickness-dependent change of the majority carrier type in $Mo_5N_6$, which is attributed to electron donation from surface -NH termination groups that becomes increasingly important at reduced thickness.

**Acknowledgment**

This material is based upon work supported by the U.S. Department of Energy (DOE), Office of Science, Basic Energy Science (BES) under Award Number DE-SC0021064 and DE-SC0026296 (X.L. and H.G.), while fabrication in the glovebox was supported by grant DE-SC0018675. X.L. acknowledges the membership of the Photonics Center at Boston University. H. G. acknowledges the support of BUnano fellowship from Boston University Nanotechnology Innovation Center. Work done by X.L. is also supported by the National Science Foundation (NSF) under Grant No. 1945364 and No. 2216008. X.Q. acknowledges the support from the U.S. Air Force Office of Scientific Research under Grant No. FA9550-24-1-0207. Y.C.L. acknowledges the partial support from Texas A&M University, and the U.S. National Science Foundation (NSF), the Future of Semiconductor 2 (FuSe2) Program, under award number DMR-2425545. The transport measurements done in this work utilizes the low-temperature transport facilities provided by the Penn State Materials Research Science and Engineering Center (MRSEC).

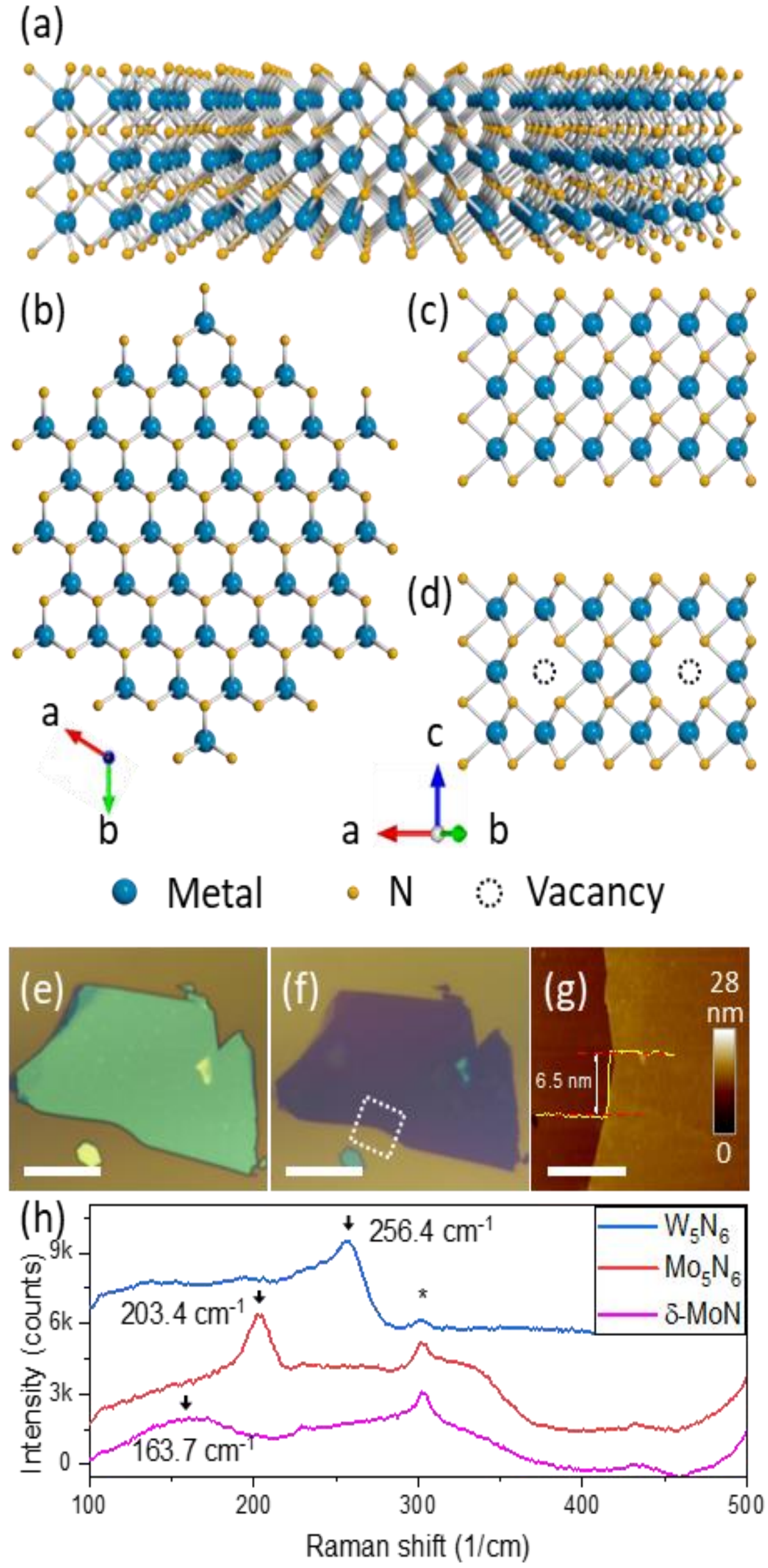

**Fig. 1 Crystal structure and preliminary characterizations of TMNs.** (a) Perspective view of the TMNs crystal. (b) Top view of a-b plane. (c, d) Side view of cation-vacancy-free TMNs (c) and cation-vacancy-rich TMNs (d). (e, f) Typical optical image of pristine $MoS_2$ (e) and converted $Mo_5N_6$ (f) flakes. (g) AFM topography measurement of the converted $Mo_5N_6$ flake. The measured area is labeled by dashed line in (f). Scale bars are 10 μm in (e) and (f), and 2 μm in (g). (h) Typical Raman spectra of different TMNs.

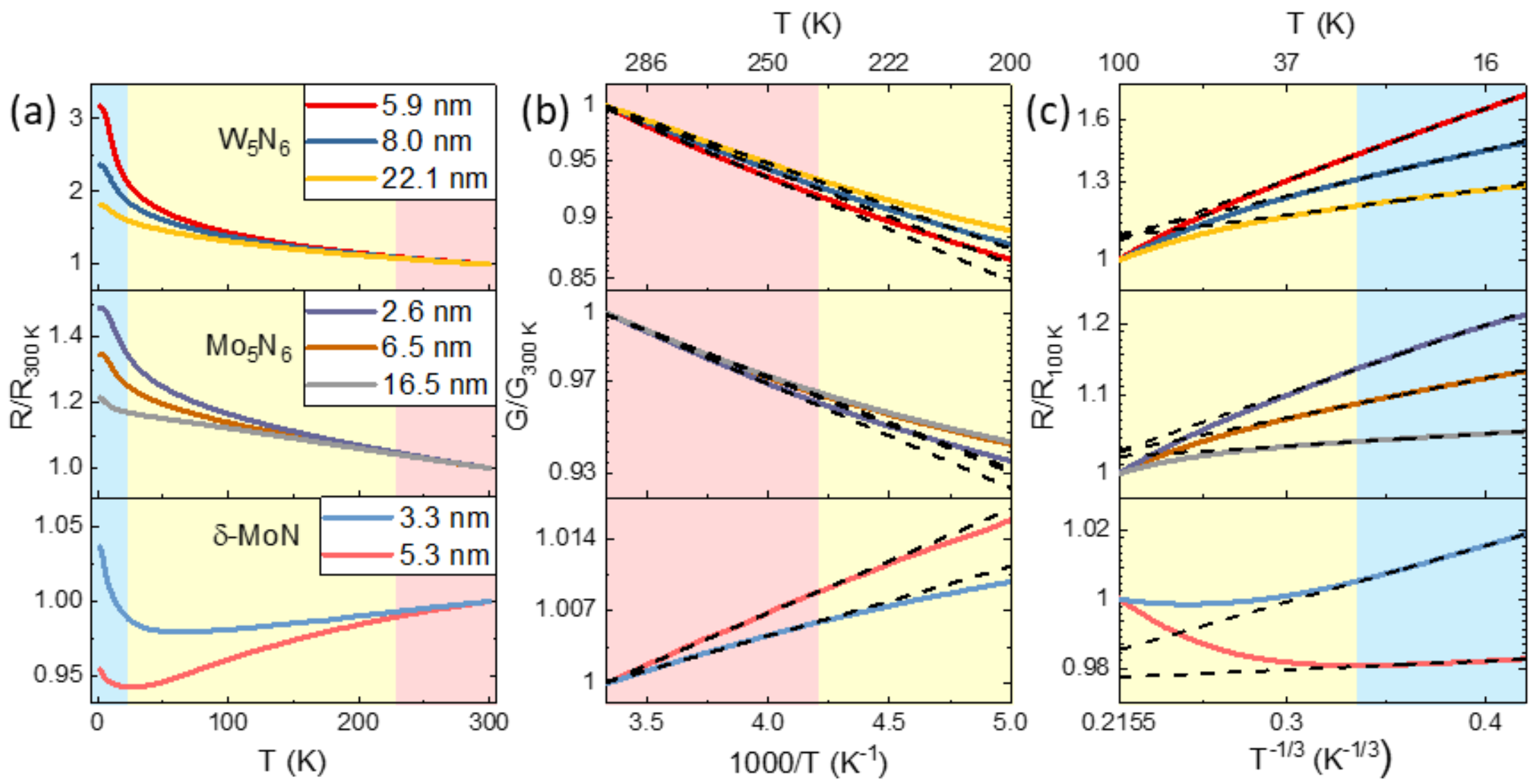

**Fig. 2 Temperature-dependent resistance measurements of TMNs.** (a) R-T plot in the range of 4-300 K. (b) G-1000/T plot in the range of 200-300 K. (c) R-$T^{-1/3}$ plot in the range of 10-100 K. Dash lines are fitting results of corresponding models discussed in the text. The color labels in (b) and (c) are inherited from (a).

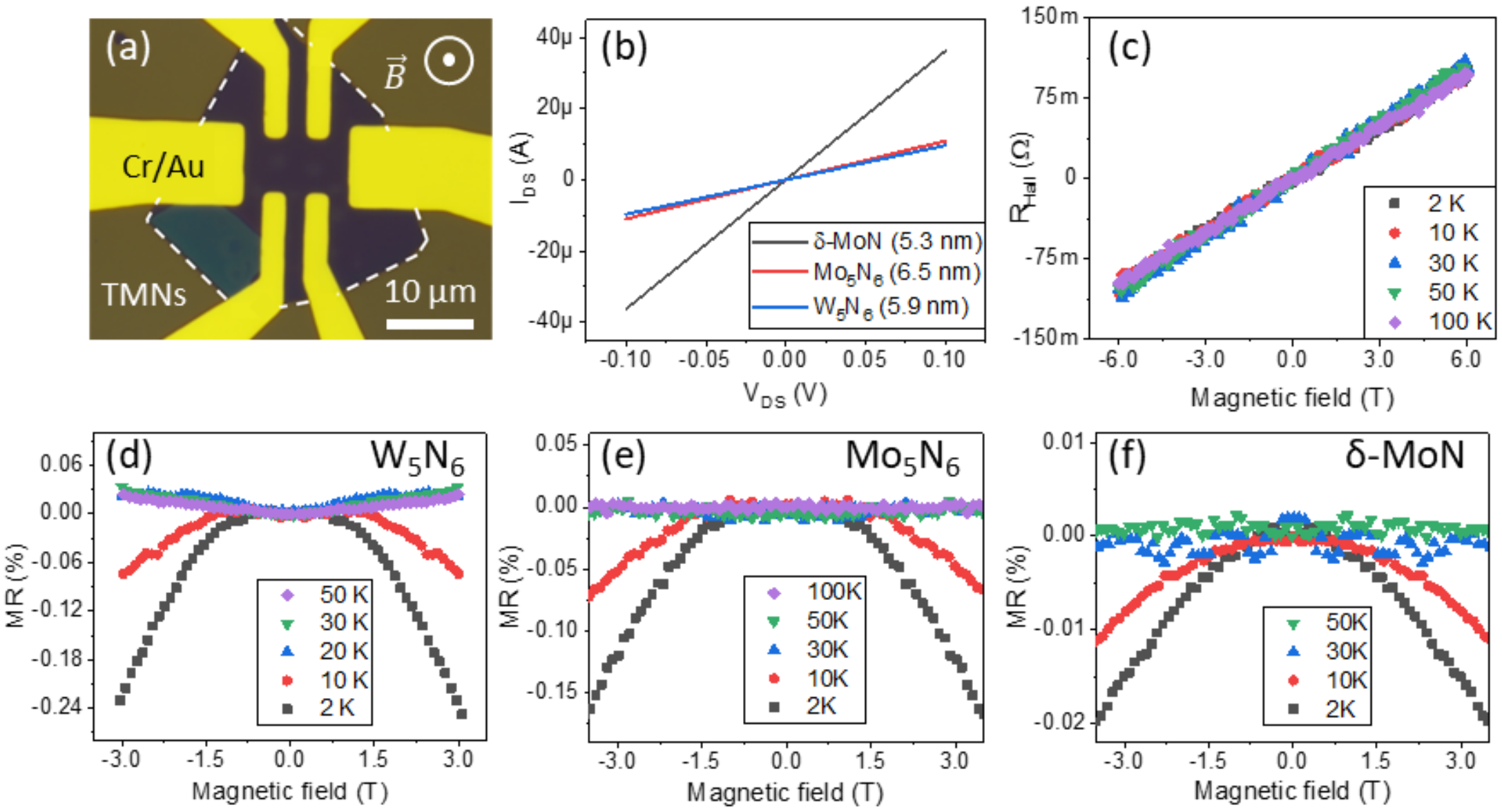

**Fig. 3 Device geometry and results of Hall measurement.** (a) Typical optical image of the sample with electrodes of Hall bar geometry. (b) I-V curve of different TMNs. (c) Typical $R_{hall}$-B results after assymetrization. (d-f) Typical magnetoresistance-B results of $W_5N_6$ (d), $Mo_5N_6$ (e) and δ-MoN (f) after symmetrization.

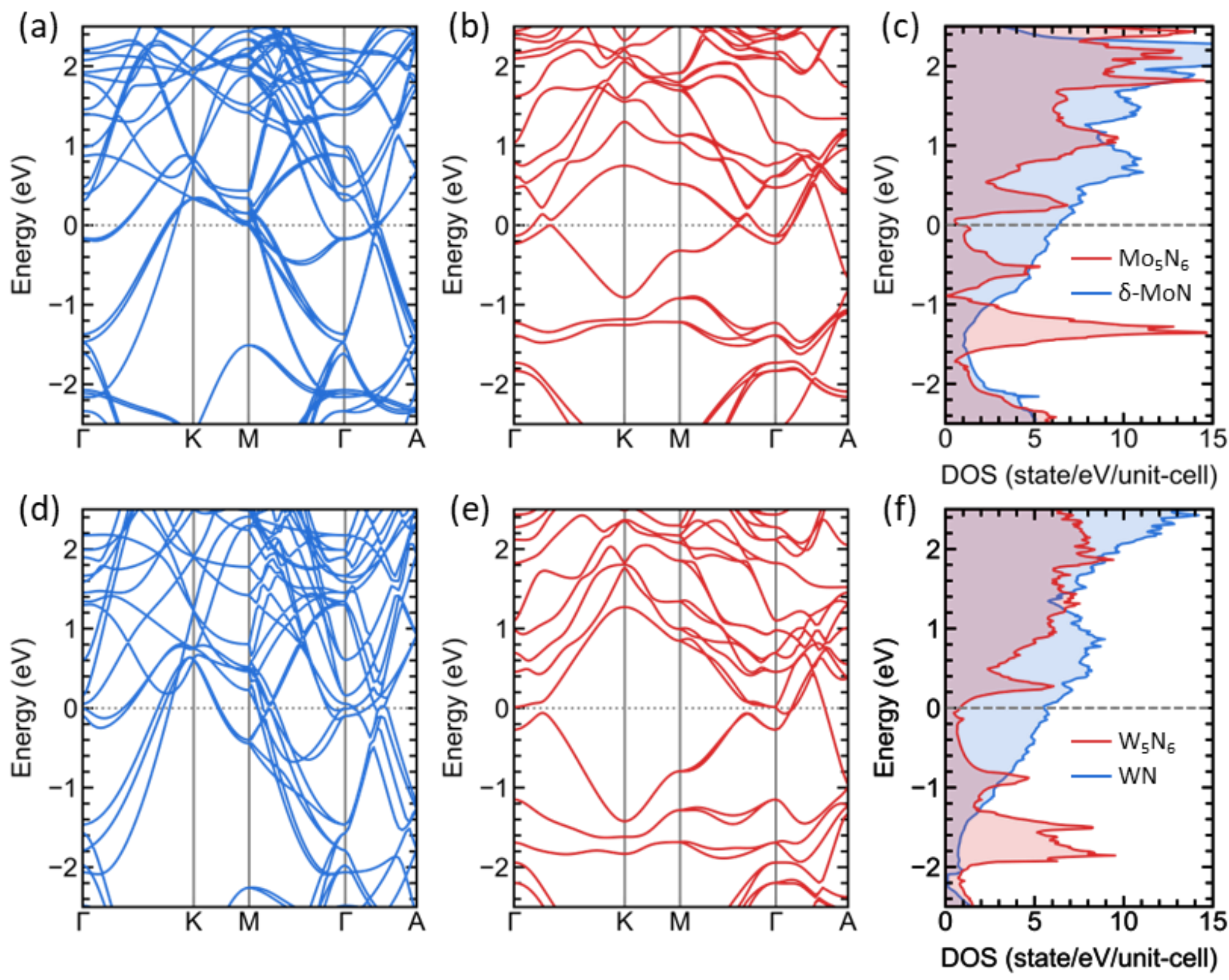

**Fig. 4 Band structure calculation of metal nitrides with and without cation vacancies.** (a, b) Energy band structure calculated based on δ-MoN (a) and $Mo_5N_6$ (b). (c) DOS comparison of δ-MoN and $Mo_5N_6$. (d, e) Energy band structure calculated based on WN (d) and $W_5N_6$ (e). (f) DOS comparison of WN and $W_5N_6$.

**Table 1 Electrical properties of TMNs extracted from I-V and Hall measurements.**

| Material | Thickness [nm] | $\sigma_{300\,K}$ [S/cm] | $n_{2D}$ [$10^{16}$ cm$^{-2}$] | $n_{3D}$ [$10^{22}$ cm$^{-3}$] | $\mu_{Hall}$ [cm$^2$ V$^{-1}$ s$^{-1}$] | Carrier type |
|---|---|---|---|---|---|---|
| $W_5N_6$ | 5.9 | 421.0 | 1.34 | 2.27 | 0.098 | n |
| | 8.0 | 312.4 | 0.57 | 0.71 | 0.23 | n |
| | 22.1 | 268.7 | 3.24 | 1.47 | 0.096 | p |
| $Mo_5N_6$ | 2.6 | 267.1 | 0.27 | 1.04 | 0.14 | n |
| | 6.5 | 426.2 | 1.58 | 2.43 | 0.092 | n |
| | 16.5 | 326.7 | 3.18 | 1.93 | 0.089 | p |
| δ-MoN | 3.3 | 3858.6 | 26.35 | 79.81 | 0.025 | n |
| | 5.3 | 3300.3 | 17.76 | 33.50 | 0.052 | p |

Note:

$\sigma_{300\,K}$ is electrical conductivity measured at 300 K. $n_{3D}$ is the 3D carrier concentration extracted from $R_{Hall}$ and considering the geometry factors. $\mu_{Hall}$ is the Hall mobility of charge carriers extracted from $R_{Hall}$. Carrier type is determined by the sign of slope of the $R_{Hall}$-B plot.

**Supplementary Materials for**

**Suppression of Metallic Transport in Nitrogen-rich Two-Dimensional Transition Metal Nitrides**

*Hongze Gao[1], Da Zhou[2], Nguyen Tuan Hung[3], Chengdong Wang[4], Zifan Wang[1], Ruiqi Lu[5], Yuxuan Cosmi Lin[4,6], Jun Cao[1], Michael Geiwitz[7], Gabriel Natale[7], Kenneth S. Burch[7], Xiaofeng Qian[4], Riichiro Saito[8,9], Mauricio Terrones[2,10,11], and Xi Ling[1,5,*]*

[1]Department of Chemistry, Boston University

590 Commonwealth Ave., Boston, MA, 02215, U.S.

[2]Department of Physics, the Pennsylvania State University

104 Davey Laboratory, University Park, PA, 16802, U.S.

[3]Department of Materials Science and Engineering, National Taiwan University

Taipei 10617, Taiwan

[4]Department of Materials Science and Engineering, Texas A&M University

575 Ross St., College Station, TX, 77843, U.S.

[5]Division of Materials Science and Engineering, Boston University

15 St Mary's St., Boston, MA, 02215, U.S.

[6]Department of Electrical and Computer Engineering, Texas A&M University

188 Bizzell St, College Station, TX 77843, U.S.

[7]Department of Physics, Boston College

140 Commonwealth Ave., Chestnut Hill, MA, 02467, U.S.

[8]Department of Physics, Tohoku University

Sendai 980-8578, Japan

[9]Department of Physics, National Taiwan Normal University

Taipei 11677, Taiwan

[10]Department of Chemistry, the Pennsylvania State University

University Park, PA, 16802, U.S.

[11]Department of Materials Science and Engineering, the Pennsylvania State University

University Park, PA, 16802, U.S.

*Corresponding author: xiling@bu.edu

In the following sections, we provide the supplementary figures to the article:

Fig. S1 Optical image of TMNs.

Fig. S2 Crystal structure of TMNs for simulations.

Fig. S3 Temperature coefficient of resistance of TMNs as a function of temperature.

Fig. S4 Activation energy (Ea) of Mo5N6 and W5N6 extracted from Arrhenius-like fitting of the G-1/T plot.

Fig. S5 $R_{hall}$-B plot of TMNs at high temperature (> 100 K). The data shown are acquired on 6.5-nm thick Mo5N6 at 150 K.

Fig. S6 Surface termination group of ultrathin $Mo_5N_6$.

Table S1 Calculated carrier density of TMNs

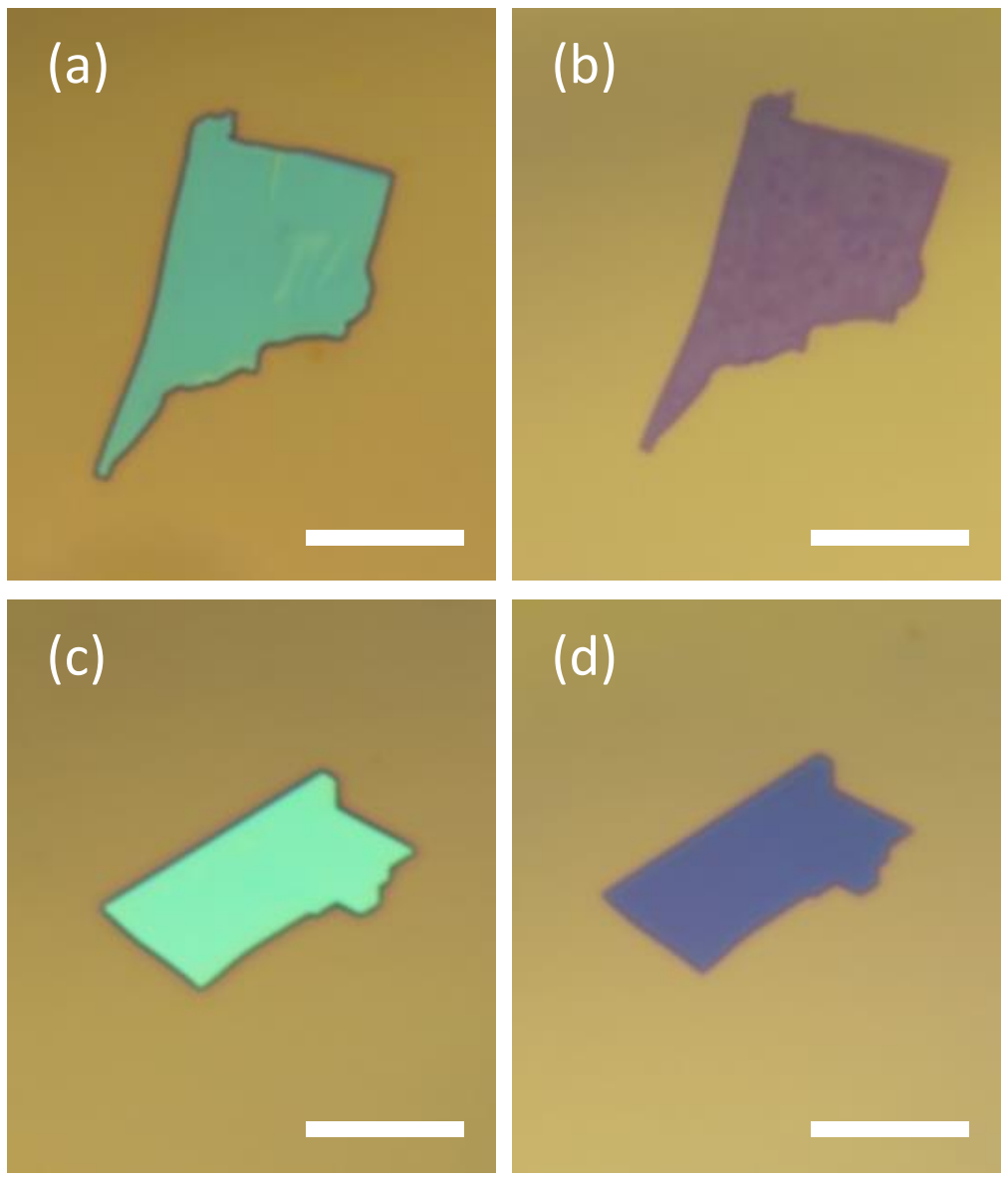

**Fig. S1** Optical image of TMNs. (a) $MoS_2$ precursor flake. (b) δ-MoN converted from $MoS_2$. (c) $WSe_2$ precursor flake. (d) $W_5N_6$ converted from $WSe_2$. All scale bars are 10 μm.

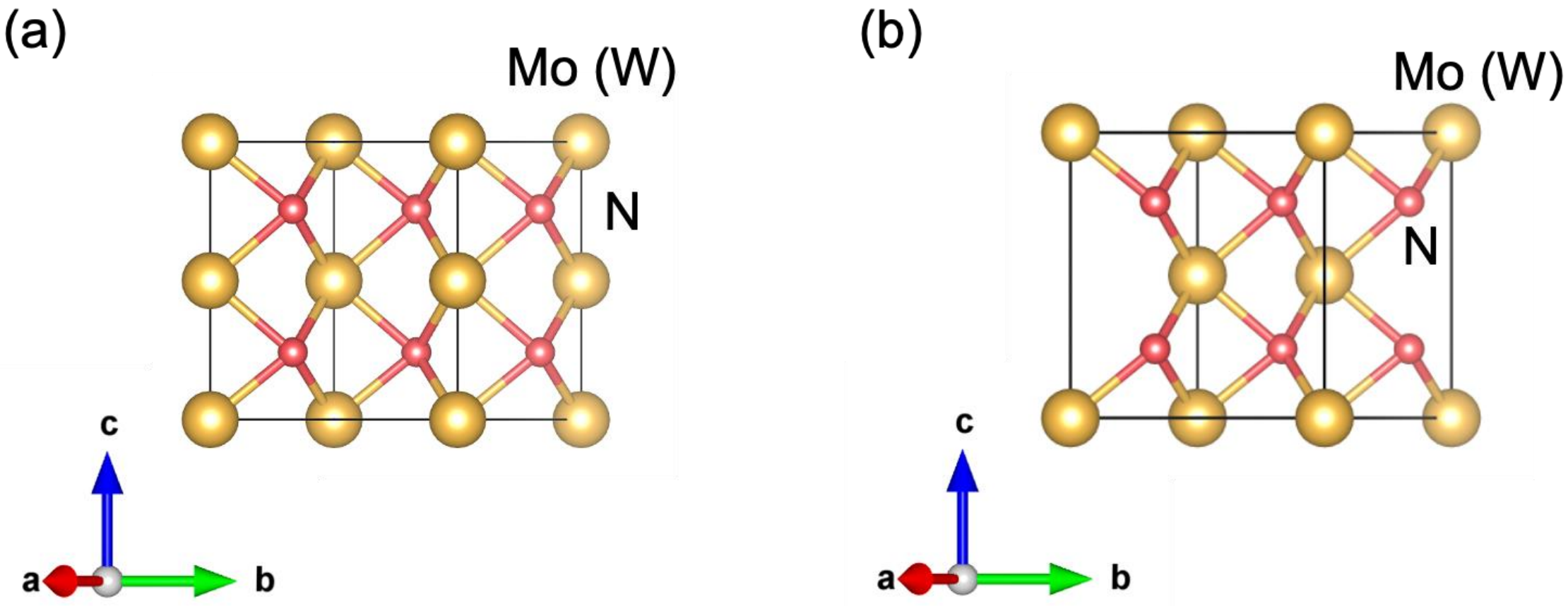

**Fig. S2** Crystal structures of TMNs for simulations. (a) Unit-cell structure of close-packed molybdenum nitride (δ-MoN) or tungsten nitride (WN). (b) Unit cell of close-packed molybdenum nitride with periodic Mo vacancies ($Mo_5N_6$) or tungsten nitride with periodic W vacancies ($W_5N_6$).

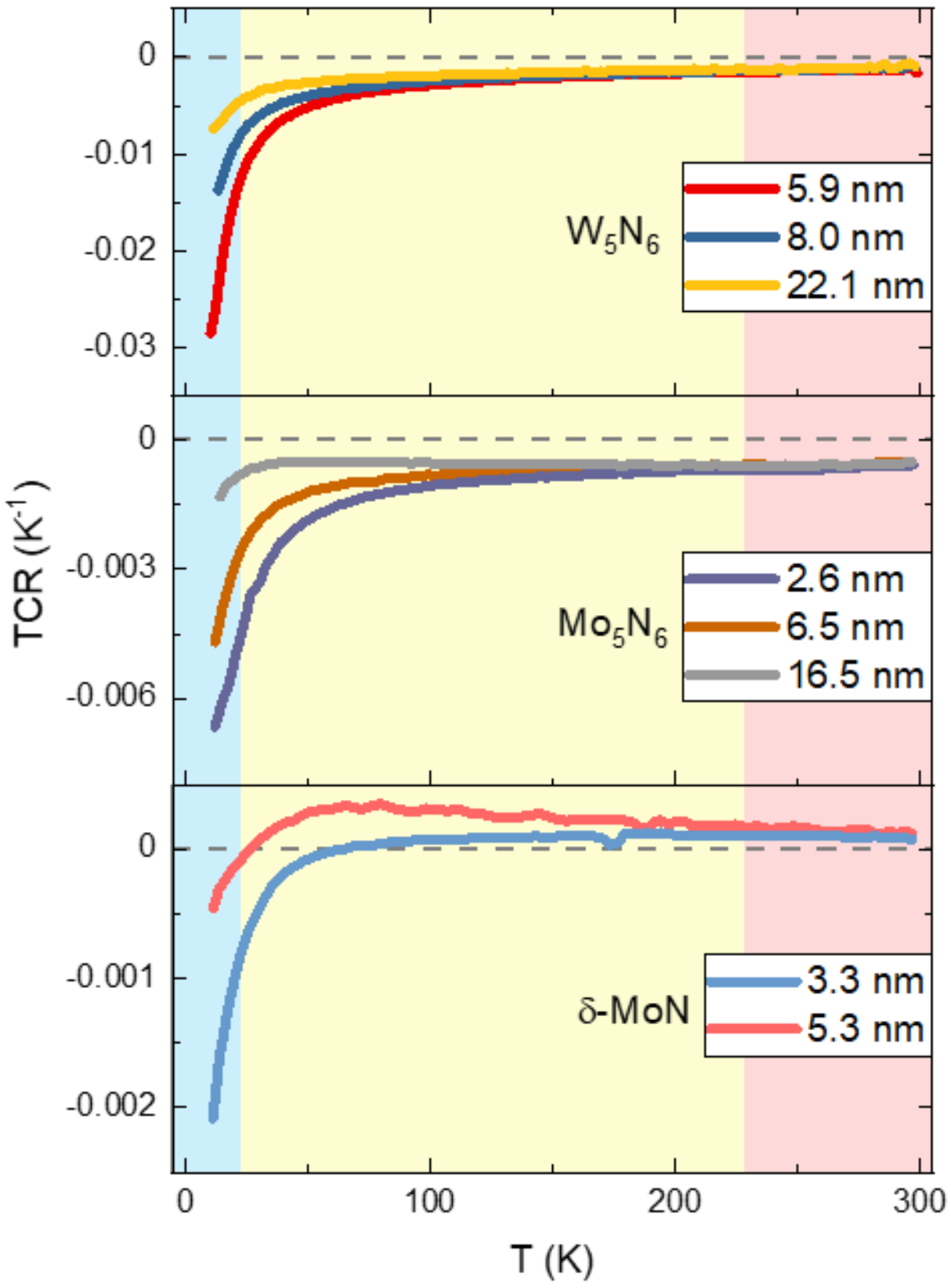

**Fig. S3** TCRs of TMNs with various thicknesses as a function of temperature.

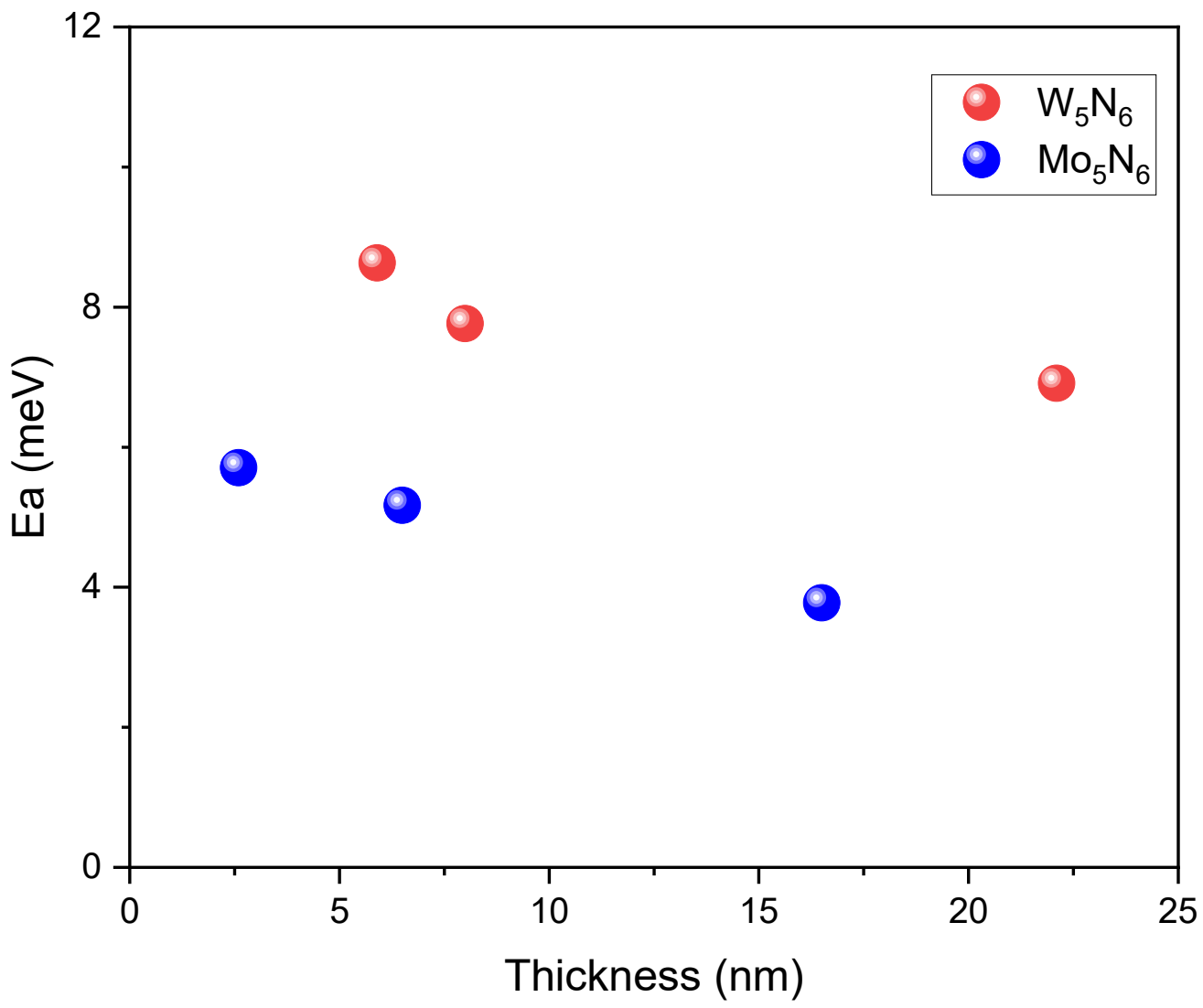

**Fig. S4** Activation energy ($E_a$) of $Mo_5N_6$ and $W_5N_6$ extracted from Arrhenius-like fitting of the G-1/T plot.

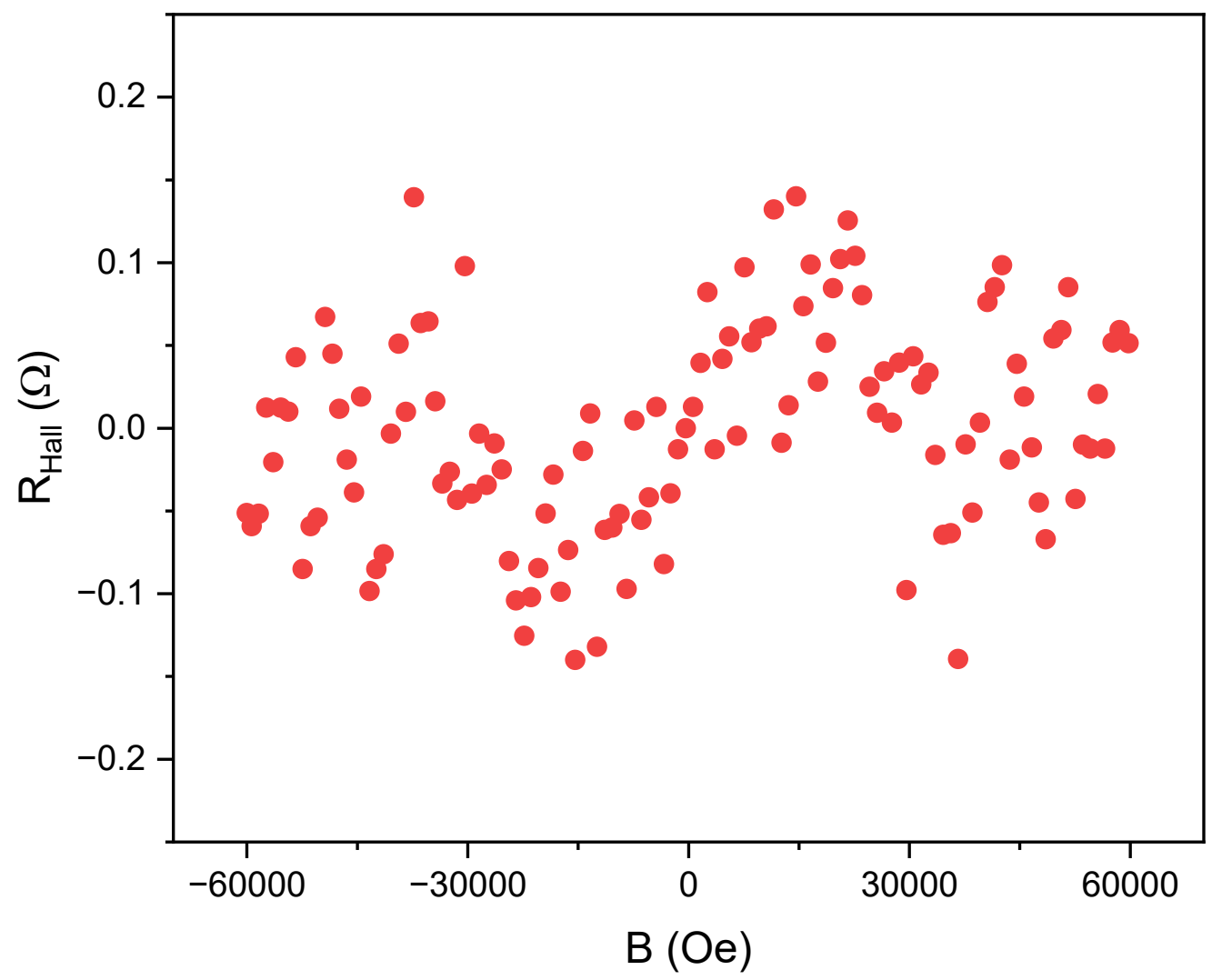

**Fig. S5 $R_{hall}$-B plot of TMNs at high temperature (> 100 K).** The data shown are acquired on 6.5-nm thick $Mo_5N_6$ at 150 K.

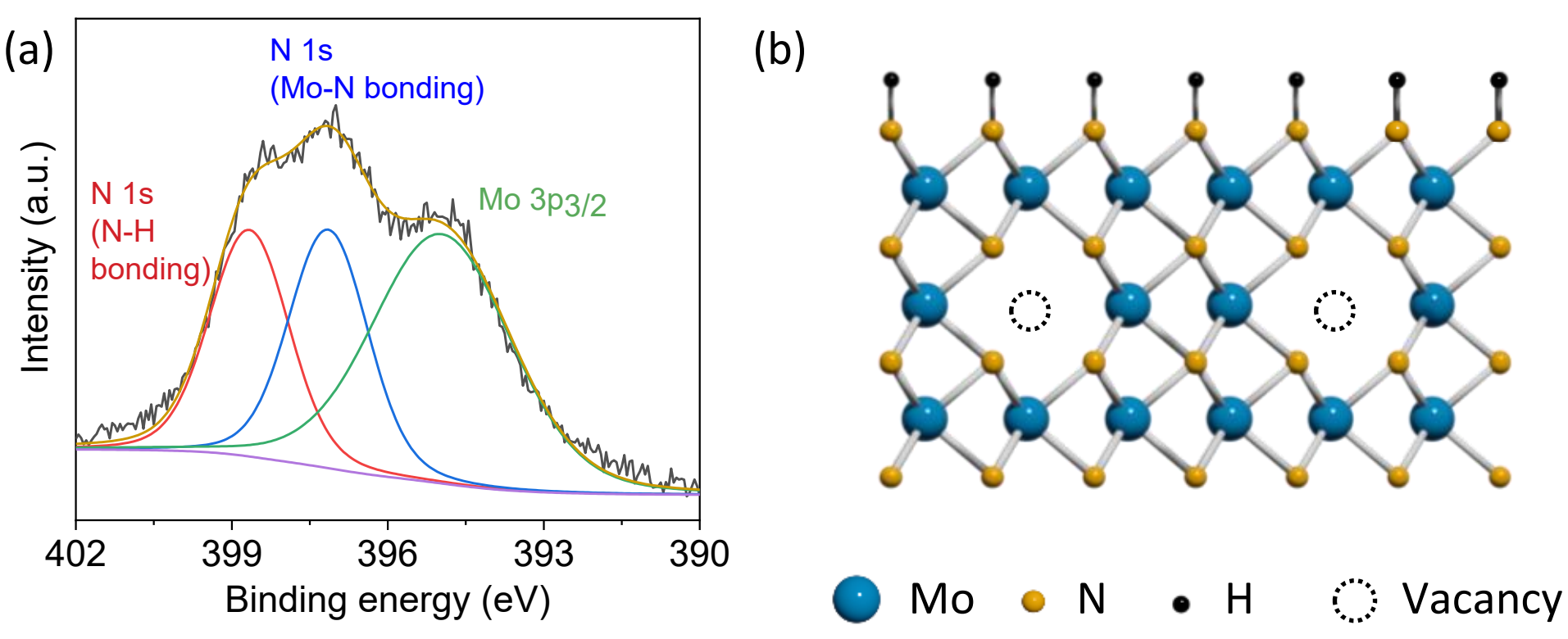

**Fig. S6** Surface termination group of ultrathin $Mo_5N_6$. (a) XPS spectrum measured on a 2-nm thick $Mo_5N_6$ flake. (b) Schematic illustration of the hydrogen-terminated surface of $Mo_5N_6$ crystal.

**Table S1 Calculated carrier density of TMNs**

| Material | Electron density [$10^{21}$ $cm^{-3}$] | Hole density [$10^{21}$ $cm^{-3}$] |
|---|---|---|
| Bulk $Mo_5N_6$ | 0.60 | 1.90 |
| 2D $Mo_5N_6$ with H terminates | 1.60 | 2.30 |
| Bulk $W_5N_6$ | 3.50 | 1.00 |
| Bulk δ-MoN | 8.50 | 6.70 |